\documentstyle[12pt]{article}
\input epsf
\begin{document}
\title{
\sc              
The single-particle density of states
and a resonance in the Aharonov-Bohm potential
\vspace*{0.3cm}}
\author{\sc Alexander Moroz\thanks{e-mail address :
{\tt am@th.ph.bham.ac.uk}}
\vspace*{0.3cm}}
\date{
\protect\normalsize
\it Division de Physique Th\'{e}orique\thanks{Unit\'{e} de 
Recherche des Universit\'{e}s Paris XI et Paris VI
associ\'{e}e au CNRS}, Institut de Physique Nucl\'{e}aire,\\
\it Universit\'{e} Paris-Sud, F-91 406 Orsay Cedex, France\\
and\\
${}$\thanks{Present address}\,School of Physics and Space 
Research, University of Birmingham,
Edgbaston, Birmingham B15 2TT, U. K.
}
\maketitle
\begin{center}
{\large\sc abstract}
\end{center}
The single-particle densitity of states (DOS) for the Pauli and the
Schr\"{o}dinger Hamiltonians in the presence of an Aharonov-Bohm 
potential is calculated for different values of the particle magnetic 
moment.
The DOS is a symmetric and periodic function of the flux.
The Krein-Friedel formula  can be applied
to this long-ranged potential when regularized with the zeta function.
We have found that whenever a bound state is present in the spectrum it
is always accompanied by a resonance.
The shape of the resonance is not of the Breit-Wigner type.
The differential scattering cross section is asymmetric if a bound state
is present and gives rise to the Hall effect. As an application, 
propagation of electrons in a dilute vortex limit is considered
and the Hall resistivity is calculated.

\vspace*{0.3cm}

{\footnotesize
\noindent PACS numbers :  03.65.Bz, 03-65.Nk, 05.30.-d, 73.50.-h}

\vspace{1.5cm}

\begin{center}
{\bf (Mod. Phys. Lett. B 9, 1407-1417 (1995))}
\end{center}
\thispagestyle{empty}
\baselineskip 20pt
\newpage
\setcounter{page}{1}
\section{Introduction}
\noindent
In this letter, nonrelativistic physics described by the 
Schr\"{o}dinger and the Pauli equations is considered in the 
presence of an Aharonov-Bohm (AB) potential ${\bf A}(r)$ \cite{AB}.
We shall use the regular  radial gauge, in which
\begin{equation}
A_r=0, \hspace*{1cm}A_\varphi=
\frac{\Phi}{2\pi r}=\frac{\alpha}{2\pi r}\,\Phi_0.
\label{abpot}
\end{equation} 
Usually, $\Phi=\alpha\,\Phi_0$ is the total flux through the flux tube
and $\alpha\geq 0$ is the total flux $\Phi$ in the units
of the flux quantum $\Phi_0$, $\Phi_0=hc/|e|$.
However, the AB potential can be considered  in a more general
sense, since, formally, the same potential (of nonmagnetic origin)
is generated around a cosmic string.
The parameter $\Phi$ is then $1/Q_{Higgs}$, $\alpha=e/Q_{Higgs}$, 
and $\Phi_0=2\pi/e$ (in the units $\hbar=c=1$) with
$e$ and $Q_{Higgs}$ being respectively the charge of a test particle
and the charge of the Higgs particle \cite{AW}.
In what follows $\alpha$ will be written as $\alpha=n+\eta$, where
$n=[\alpha]$ is the nearest  integer {\em smaller} than 
or equal to $\alpha$ and $\eta$ being the fractional part.
The case of a nonsingular flux tube of finite radius $R$
 will be discussed, too,
as it is important from the experimental point of view. Indeed,
flux tubes realized in experiments such as vortices
in superconductor of type II are never of zero radius.
Our main results are: 

\begin{itemize}
\item the validity of the  Krein-Friedel formula \cite{F,AMB}
for the density of states (DOS) is for the first time established 
for a singular potential
and  the change  $\triangle\rho_\alpha(E)$
over all space of the DOS induced  by the AB potential is calculated;
\item a resonance is  predicted to occur whenever a bound state
is present in the spectrum;
\item in contrast to zero modes \cite{AC}, the number of bound states
 does depend on the regularization of the interior of a flux tube;
\item  in the presence of a bound state, the differential scattering 
cross section is asymmetric  and the Hall effect occurs.
\end{itemize}
Details of our calculations and complete proofs are given elsewhere
\cite{AM20,AM30,AM89}. Here, main ideas  are presented
and  some of the proofs are outlined.

The  Krein-Friedel formula  (\ref{krein}) gives the DOS as the sum over 
phase shifts and thereby relates the DOS directly 
to the scattering properties. 
It is therefore very useful to have its extension in the direction of 
singular (especially Coulomb) potentials.
Here the  Krein-Friedel formula is applied directly 
and consistency with previous
results is shown. For example, in the particular case
without bound states we confirm the anticipation of 
Comtet, Georgelin, and Ouvry \cite{CGO} that the
change of the DOS is concentrated at zero energy.
The DOS provides an important link between different physical quantities:
the partition function, virial coefficients, effective action, 
and in the relativistic case between
induced fermion number and anomaly \cite{NS,AM89}. We have already 
used its knowledge to
calculate the persistent current of free electrons induced  in the plane
by the AB potential \cite{CMO1}.

\begin{figure}
\centerline{\epsfxsize=8cm \epsfbox{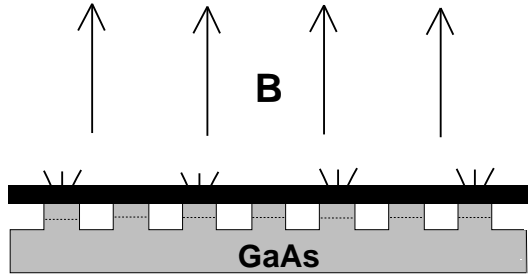}}
\caption{Black layer is a superconductor of type II put on top
of the heterostructure containing the two-dimensional electron gas
(dotted region). When this sample is put in a homogeneous magnetic
field, magnetic field  penetrates the superconductor
in Abrikosov vortices. Therefore, electrons from the heterostructure
do not move in the homogeneous magnetic field but in the field
of a (penetrable) flux tube.}
\label{figrssc}
\end{figure}

The discovery of a resonance was quite unexpected.
 Rather suprisingly, the shape of the resonance 
[can be read from Eq. (\ref{resshape})] is {\em not} of the 
Breit-Wigner form. Since the latter is a direct
consequence of analyticity, it poses an interesting question on
the analytic structure of scattering amplitudes for singular
potentials. 
The resonance can have a profound influence on the transport properties
of electrons in an experimental set-up where electrons
can penetrate the interior of the
flux tube provided that the latter is prepared is such a way that
its interior is not isolated from the system under consideration.
A realistic physical realization of the penetrable
flux tube is that suggested originally by
Rammer and Shelankov \cite{RS} and later realized experimentally
by Bending, Klitzing, and Ploog \cite{BKP}, i.\ e., to put
a type II superconducting gate on top
of the heterostructure containing the two-dimensional
electron gas (2DEG) (see Fig.\ \ref{figrssc}). 
When a magnetic field is switched on the conventional
superconductor is penetrated by vortices of flux with $\alpha=1/2$.
Therefore, electrons from the heterostructure do not move in
the homogeneous magnetic field but in the field of a 
penetrable flux tube.

By considering the case of a regular flux tube with a finite
radius $R$ we shall show that a bound state can only occur
if the gyromagnetic 
ratio $g_m$ is  anomalous and greater than  two.
If $g_m$ equals exactly to two, then according to Aharonov-Casher 
theorem \cite{AC} zero modes occur.
If $g_m<2$,  the coupling with magnetic field is not sufficiently
strong enough to form neither zero modes nor bound states.
In the region $g_m>2$, i.e., exactly where the gyromagnetic ratio
of electron ($g_m=2.00232$) lies, the coupling with  the
magnetic field is enhanced and zero modes turn out to be bound states.
However, the number of bound state is generally higher than the number
of zero modes and the number of bound states does not only depend on 
the total flux but also on the
energy of the magnetic field. 

The differential scattering cross section is a periodic
function of the flux $\alpha$ and asymmetric with respect to
$\varphi\rightarrow -\varphi$, where $\varphi$ is the scattering 
angle \cite{AM30}.
The asymmetry of the differential scattering cross section
 is easy to understand as for $\alpha\geq 0$ bound states
occur only for $l\leq 0$, $l$ being the orbital angular momentum.
The asymmetry of the differential scattering cross section has direct
experimental consequences since it leads to the Hall effect.
The Hall resistivity is calculated in the dilute vortex limit.

One has the unitary equivalence between a spin $1/2$ charged 
particle in
a 2D magnetic field and a spin $1/2$ neutral particle with an 
anomalous magnetic moment in a 2D electric field \cite{OO} 
and our results apply to this case as well.

\section{The Pauli and the Schr\"{o}dinger Hamiltonian
in the Aharonov-Bohm potential and scattering phase shifts}
Let us consider the Pauli Hamiltonian,
\begin{equation}
H= \frac{({\bf p}-\frac{e}{\hbar c}{\bf A})^2}{2m}-
\hat{\mbox{\boldmath$\mu$}}\cdot{\bf B},
\label{pauli}
\end{equation}
where $\hat{\mbox{\boldmath$\mu$}}=\mu\hat{{\bf s}}/s$ is the magnetic
moment operator, $\hat{{\bf s}}$ is the spin operator,
and $s$ is the magnitude of the particle spin.
For electron $\mu_e =-g_m|e|\hbar/4mc=-\mu_Bg_m/2$, $\mu_B$ being the 
Bohr magneton,  and $g_m$ is the gyromagnetic ratio that characterizes 
the strength of the magnetic moment \cite{Land}.
By separating the variables, assuming $e=-|e|$, $H$
is written as a direct sum, $H=\oplus_l H_l$, of channel 
radial Hamiltonians $H_l$ in the Hilbert space
$L^2[(0,\infty), rdr]$ \cite{AB,R},
\begin{equation}
H_l=-\frac{d^2}{dr^2}-\frac{1}{r}\frac{d}{dr}+\frac{\nu^2}{r^2}
+g_m\frac{\alpha}{r} s_z\delta(r).
\label{schrham}
\end{equation}
Here $\nu=|l+\alpha|$
and $s_z=\pm 1$ is the projection of the spin on the direction 
of the flux tube \cite{AB,R}.
The Schr\"{o}dinger equation is recovered upon setting $s_z=0$.
For positive (negative) energies the eigenvalue equation in the 
$l$-th channel 
reduces to the (modified) Bessel equation of the order 
$\nu=|l+\alpha|$,
\begin{equation}
H_l\,\psi_l=k^2\psi_l
\label{eigen}
\end{equation}
with $k=\sqrt{2mE}/\hbar$. 
In case of the impenetrable flux tube the spectra 
of both the Pauli and the Schr\"{o}dinger equations are identical.
There are neither zero modes nor bound states in this case
\cite{AM30}.
The boundary condition selects only regular solutions at the 
origin and the ``spectrum"  is given by
\begin{equation}
\psi_l(r,\varphi)=J_{|l+\alpha|}(kr)e^{il\varphi}.
\label{regpsi}
\end{equation}
Phase shifts \cite{AB},
\begin{equation}
\delta_l = \frac{1}{2}\pi(|l|-|l+\alpha|),
\label{convshift}
\end{equation}
are in general singular: they do not decay to zero
in the limit $E\rightarrow \infty$.

In general case, Hamiltonians $H_l$ for which $|l+\alpha|<1$
admit a one-parametric family of self-adjoint extensions \cite{R,RR}.
They correspond to different physics inside the flux tube.
The situation will be  considered when bound states 
\begin{equation}
B_l(r,\varphi)=K_{|l+\alpha|}(\kappa_l r)e^{il\varphi}
\label{bounst}
\end{equation}
of energy $E_l=-(\hbar^2/2m)\kappa_l^2$ are present 
in the  $l=-n,-n-1$ channels, with $n=[\alpha]$.
In the presence of the bound states the scattering states 
(\ref{regpsi}) in these channels have to be modified. They become
\begin{equation}
\psi_l(r,\varphi)=
\left[J_{|l+\alpha|}(kr)-A_lJ_{-|l+\alpha|}
(kr)\right] e^{il\varphi}.
\label{psising}
\end{equation}
This is because $H_l$ has  necessarily to be a symmetric operator
what already determines $A_l$ to be 
\begin{equation}
A_l=(k/\kappa_l)^{2\nu}, 
\end{equation}
i.e., {\em energy dependent} \cite{AM30}.
The radial part of the general solution (\ref{psising}) behaves 
for $(r\rightarrow\infty)$ as
\begin{equation}
R_l(r)\sim \mbox{const}\left(e^{-ikr}+\frac{1-A_le^{i\pi|l+\alpha|}}
{1-A_le^{-i\pi|l+\alpha|}}e^{-i\pi(|l+\alpha|+1/2)}
e^{ikr}\right)\nonumber
\end{equation}
Therefore,
\begin{equation}
\delta_l=\frac{1}{2}\pi(|l|-|l+\alpha|)+\arctan\left(\frac{\sin(|l+
\alpha|\pi)}{\cos(|l+\alpha|\pi) -A_l^{-1}}\right),
\label{shift}
\end{equation}
which determines S matrices, S${}_l=e^{2i\delta_l}$,
in these two channels.

Note that a bound state has the most profound influence on phase
shifts in the limit $E_b\uparrow 0$. In this limit
\begin{equation}
\delta_l \rightarrow 
\frac{1}{2}\pi(|l|+|l+\alpha|)
\end{equation}
and the phase-shift flip occurs (cf. \cite{Hag}).
On contrary, in the limit $E_b\downarrow -\infty$,
\begin{equation}
\delta_l\rightarrow \frac{1}{2}\pi(|l|-|l+\alpha|).
\end{equation}

\section{The Krein-Friedel formula and the DOS}
The DOS in the presence of the AB potential is defined to be
\begin{equation}
\rho_\alpha(E)\equiv -\frac{1}{\pi}\mbox{ImTr}\,G_\alpha
({\bf x},{\bf x},E+i\epsilon),
\label{resf}
\end{equation}
where  $G_\alpha({\bf x},{\bf y},E+i\epsilon)$ the resolvent
(the Green function) of $H$.
The integrated density of states $N_{\alpha}(E)$ is then as usual
given by
\begin{equation}
N_\alpha(E)\equiv \int_{-\infty}^E \rho_\alpha(E')\,dE'.
\end{equation}
The DOS in  two dimensions when all interactions are
switched-off is $\rho_0(E)=(m/2\pi\hbar^2)V$, with
$V=\int\,d^2{\bf r}$ being the (infinite) volume.
To calculate the change of the  {\em integrated}
density of states (IDOS) in the whole space
we shall make use of the Krein-Friedel formula \cite{F}
that gives the change $\triangle N_\alpha(E)$
of the IDOS induced by the
the presence of a scatterer of a finite range 
directly by summing over phase shifts,
\begin{equation}
\triangle N_\alpha(E)\equiv 
N_\alpha(E)-N_o(E)=\frac{1}{\pi}\sum_l \delta_l(E)=
(2\pi i)^{-1}\ln\det\mbox{S},
\label{krein}
\end{equation}
with S the total on-shell S-matrix.
The fact that phase shifts can be rather easily calculated
without any care of the proper normalization of wave functions
greatly facilitates the calculation. Moreover, by means of the
Krein-Friedel formula it is rather easy to calculate the
change of the IDOS for all possible self-adjoint extensions.
In the case of the long-ranged AB potential we have found that
the Krein-Friedel formula 
when combined with the $\zeta$-function regularization
can  still be used despite the fact that phase shifts (\ref{convshift})
are in general singular \cite{AM20,AM30}.
In the absence of bound states,
\begin{eqnarray}
\lefteqn{\ln\det\mbox{S}=\sum_{l=-\infty}^\infty 2i\delta_l
= i\pi\sum_{l=-\infty}^\infty (|l|-|l+\alpha|)}\nonumber\\
&&=
i\pi\left.\left[ 2\sum_{l=1}^\infty l^{-s}-\sum_{l=0}^\infty (l+\eta)^{-s}
-\sum_{l=1}^\infty (l-\eta)^{-s}\right]\right|_{s=-1}
\nonumber  \\
&&
 =i\pi\left.\left[2\zeta_R(s)-\zeta_H(s,\eta)-\zeta_H(s,1-\eta)\right]
\right|_{s=-1}=
-i\pi\eta(1-\eta),
\end{eqnarray}
where $\zeta_R$ and $\zeta_H$ are the Riemann and the Hurwitz 
$\zeta$-function.
Thus, using  (\ref{krein}),  the change of the DOS is
\begin{equation}
\triangle\rho_\alpha(E)=\rho_\alpha(E)
-\rho_0(E)=- \frac{1}{2}\,\eta(1-\eta)\,\delta(E),
\label{denreg}
\end{equation}
and $\triangle\rho_\alpha(E)$ is only the function 
of a distance from the nearest integer.

In the presence of bound states,  the contribution
of scattering states to $\triangle N_\alpha$ for $E\geq 0$ is 
\begin{eqnarray}
\triangle N_\alpha(E)= &-&\frac{1}{2}\eta(1-\eta)+\frac{1}{\pi}\arctan
\left(\frac{\sin(\eta\pi)}{\cos(\eta\pi)-(|E_{-n}|/E)^{\eta}}
\right)
\nonumber\\
&-&\frac{1}{\pi}\arctan
\left(\frac{\sin(\eta\pi)}{\cos(\eta\pi)+
(|E_{-n-1}|/E)^{(1-\eta)}}\right),
\label{intsing}
\end{eqnarray}
where $E_{-n}$ and $E_{-n-1}$ are the binding energies 
in $l=-n$ and $l=-n-1$ channels. 
By repeating the same calculation for $\alpha\leq 0$ one finds that
$\triangle N_\alpha(E)$ is a symmetric function of $\alpha$
\cite{AM30},
\begin{equation}
\triangle N_{-\alpha}(E)=\triangle N_{\alpha}(E).
\end{equation}

\subsection{The resonance}
Note that for 
$0<\eta<1/2$ the {\em resonance} appears at
\begin{equation}
E=\frac{|E_{-n}|}{\left[\cos(\eta\pi)\right]^{1/\eta}}>0.
\nonumber
\end{equation}
The phase shift 
$\delta_{-n}(E)$ (\ref{shift}) changes by $\pi$ in the direction
of increasing energy and  the integrated density of states 
(\ref{intsing}) has a sharp
increase by one.  The  profile of the resonance 
[the argument of $arctan$ in (\ref{shift})] is given by
\begin{equation}
 \frac{E^\eta \tan \eta \pi}{E^\eta - E_{res}^\eta}=
\frac{\Gamma}{E^\eta - E_{res}^\eta},
\label{resshape}
\end{equation}
where $\Gamma= E_{res}^\eta \tan \eta \pi$ is the 
width of the resonance. Note that its profile (\ref{resshape})
{\em is not} of the Breit-Wigner form (see Ref.\  \cite{La}, \& 145),
\begin{equation}
\frac{\Gamma}{E-E_{res}}\cdot
\label{bfform}
\end{equation}
For $1/2<\eta<1$ the resonance is shifted to the 
$l=-n-1$ channel.
$\eta=1/2$ is a special point since resonances occur in both channels
at infinity. Therefore the contribution
of the {\em arctan} terms in (\ref{intsing}) 
does not vanish as $E\rightarrow\infty$, but instead gives $-1$.

\section{Regularization, $R\rightarrow 0$ limit,
and the interpretation of self-adjoint extensions}
Different self-adjoint extensions correspond to different
physics inside the flux tube (see an example in Ref. \cite{RS2}, p. 144).
To identify the physics which underlines them
we have considered the situation when the AB potential
is regularized by a {\em uniform} magnetic field $B$ within the radius $R$
and satisfies the constraint
\begin{equation}
\int_\Omega B({\bf r})\, d^2 {\bf r} =\Phi =  \mbox{const}.
\label{bconst}
\end{equation}
One finds that in the absence of the magnetic moment ($g_m=0$) or any
other attractive interaction with the interior of the flux tube
the matching equation for the exterior and interior 
solutions in the $l$-th channel is
(see \cite{BV} for example),
\begin{equation}
x\frac{K_{|l+\alpha|}'(x)}{K_{|l+\alpha|}(x)}=-\alpha+|l|+\alpha
\frac{|l|+l+1+(x^2/2\alpha)}
{|l|+1}\frac{
{}_1F_1\left(\frac{|l|+l+3}{2}+(x^2/4\alpha),|l|+2,\alpha\right)}
{{}_1F_1\left(\frac{|l|+l+1}{2}+(x^2/4\alpha),|l|+1,\alpha\right)}\cdot
\label{matching}
\end{equation}
Here ${}_1F_1(a,b,c)$ is the Kummer hypergeometric function \cite{AS},
and $x_l=\kappa_l R\neq 0$. However, since the l.h.s.
decreases from $-|l+\alpha|$ to $-\infty$ as $x\rightarrow\infty$ and
the r.h.s. is always greater than $-l+|\alpha|$ one finds that 
Eq.\ (\ref{matching}) does not have a solution unless it is an
{\em attractive} potential $V(r)$ inside the flux tube,
\begin{equation}
V(r)|_{r\leq R}=-\frac{\hbar^2}{2m}\,\frac{\alpha}{R^2}\,c(R),
\label{tubpot}
\end{equation}
where $V(r)=0$ otherwise. Here, $c(R)=2(1+\varepsilon(R))$, 
$\varepsilon(R)>0$, and $\varepsilon(R)\rightarrow 0$ 
as $R\rightarrow 0$ \cite{BV,BF}. 
This amounts to changing $x^2/2\alpha$ to $ x^2/2\alpha -c/2$ 
on the r.h.s. of (\ref{matching}). 
Note that in the limit $R\rightarrow 0$
\begin{equation}
V(r)|_{r\leq R} \rightarrow -\frac{\hbar^2}{m}\frac{\alpha}{r}\delta(r).
\label{tubpol}
\end{equation}
The attractive potential can be either put in
by hand or, if the Pauli Hamiltonian is considered, as arising
from the magnetic moment coupling of the electrons 
with spin  opposite to the direction of the magnetic field $B$.
In the latter case the critical potential corresponds to the case
when gyromagnetic ratio $g_m=2$ [$\varepsilon(R)=(g_m-2)/2\equiv 0$]
(cf. Eq.\ (\ref{schrham}). Then the matching equation 
(\ref{matching}) has a solution in $x=0$ for $l=-n$. 
Since the magnetic field is not singular any more
 the Aharonov-Casher theorem \cite{AC} applies.
It is known that there are $]\alpha[-1$  {\em zero modes}
in this case, $]\alpha[$ the nearest integer {\em larger} than or 
equal to $\alpha$.
If $\alpha$ is an integer then one has exactly $n-1$ zero modes, 
if not, their number is  $n=[\alpha]$
\cite{AC}. The result  only depends on the total flux $\alpha$ and
 not on a particular distribution of a magnetic field $B$.

Whenever $g_m>2$ (and hence $\varepsilon>0$)
or $g_m=2$ with an attractive potential 
$V(r)=-\varepsilon/R^2$, $\varepsilon>0$ arbitrary small,
the bound states may occur in the spectrum in the channels
$l\leq 0$.
 They correspond to solutions $x_l>0$ of (\ref{matching}).
In other words the coupling with the interior of the flux tube
 becomes sufficiently strong for the particle to be confined
on the cyclotron orbit {\em inside} it. Note that the wave function
(\ref{bounst}) of bound state decays exponentially outside the 
flux tube. In contrast to the zero modes their number {\em does depend}
on a particular distribution of the magnetic field $B$.
Using three different regularizations, uniform, regular, and 
cylindrical one finds that the number of bound states  is less 
than or equals to  \cite{AM20,AM30,BV1}
\begin{equation}
\#_b =1 +n + \left[\alpha(g_m-2)/4\right] +
\left[\alpha(g_m+2)/4 -n\right],
\nonumber
\end{equation}                     
with $[.]$ as above.  Note that the number of bound states is
generally higher than the number of zero modes \cite{AM30}.
The bound is saturated \cite{AM20,AM30} if  the 
cylindrical shell regularization \cite{Hag}
of the AB potential is used.
The physical origin of this difference  can be understood in a
simple way. In the latter case the energy
$E_B$ of magnetic field is {\em infinite} for any $R\neq 0$ 
and in this sense the magnetic field inside the flux tube is much 
stronger than, for example,
in the homogeneous field regularization when $E_B$,
\begin{equation}
E_B=\pi B^2 R^2/2 = \Phi^2/(2\pi R^2),
\label{mgen}
\end{equation}
stays {\em finite} for any nonzero $R$.

\section{The Hall effect}
The S matrix, $s_\alpha(\varphi)$,
 in the AB potential was calculated according to
\begin{equation}
s_\alpha(\varphi) \equiv 
\frac{1}{2\pi} \sum_{l=-\infty}^\infty e^{2i\delta_l+il\varphi}.
\end{equation}
We have found that either in the absence or presence of 
bound states,
\begin{equation}
s_{-|\alpha|}(\varphi)= s_{|\alpha|}(-\varphi),
\label{sminv}
\end{equation}
under the transformation $\alpha\rightarrow -\alpha$ \cite{AM30}.

In the presence of bound states,
the differential scattering cross section for $\varphi\neq 0$
was found to be \cite{AM30}
\begin{eqnarray}
\lefteqn{
\left(\frac{d\sigma}{d\varphi}\right)(k,\varphi) =
\left(\frac{d\sigma^0}{d\varphi}\right)(k,\varphi) +
\frac{8\pi}{k}\sum_{l=-n-1}^{-n}\sin^2\triangle_{l}
}\nonumber\\
&&
\mbox{}+ \frac{4}{k}\frac{\sin(\pi\alpha)}{\sin(\varphi/2)}
\left[
\sin\triangle_{-n}\cos\left(\triangle_{-n}-\pi\alpha
+\varphi/2\right) +\sin\triangle_{-n-1}
\cos\left(\triangle_{-n-1}+\pi\alpha
-\varphi/2\right) \right],\hspace*{1cm}
\label{dbcross}
\end{eqnarray}
where
\begin{equation}
\left(\frac{d\sigma^0}{d\varphi}\right)(k,\varphi) =\frac{1}{2\pi k}
\frac{\sin^2 (\pi\alpha)}{\sin^2(\varphi/2)}
\label{dcross}
\end{equation}
is the differential scattering cross section in the absence 
of bound states \cite{R}. 
The periodicity of the differential cross section with respect to the substitution $\alpha\rightarrow \alpha \pm 1$ then follows from Eq.\
(\ref{dbcross}).
Note that in the presence of bound states, the differential
cross section becomes {\em asymmetric} with regard to 
$\varphi\rightarrow-\varphi$ (what is equivalent, with regard to
$\alpha\rightarrow -\alpha$). The origin of the asymmetry is 
easy to understand since bound states for $\alpha\geq 0$ are 
only formed in channels with $l\leq 0$. 

Now, if one considers  a random distribution of flux tubes
with the density $n_v$, then the Hall effect is induced \cite{AM30}.
The Hall resistivity $\rho_{xy}$ can be calculated in the
dilute vortex limit, i.\ e., when the multiple-scattering
effects are ignored.
The quantity that measures the fraction of the electrons moving
in a transverse direction is $\sin(\varphi)\, d\sigma(k_F,\varphi)$.
Therefore, if the density of vortices is $n_v$,
the Hall current in the dilute vortex limit
is proportional to
\begin{equation}
n_v\, \int_{-\pi}^\pi\!
d\varphi\, \sin\varphi\, \frac{d\sigma}{d\varphi}(k_F,\varphi).
\label{holik}
\end{equation}
By inverting the conductivity tensor one finds that  
the Hall resistivity, $\rho_{xy}$, is
\begin{equation}
\rho_{xy}=  \rho_H^0 \frac{k_F}{\alpha} \int_{-\pi}^\pi\!
\frac{d\varphi}{2\pi}\sin\varphi\, \frac{d\sigma}{d\varphi}(k_F,\varphi),
\label{hall}
\end{equation}
which was also obtained by Nielsen and Hedegaard \cite{NH}.
Substituting (\ref{dbcross}) to (\ref{hall}) then gives \cite{AM30}
\begin{equation}
\rho_{xy}= \frac{4n_v}{n_e}\frac{hc}{e^2}\sin(\pi\alpha)\left[
\sin\triangle_{-n} \cos(\triangle_{-n}-\pi\alpha)
+\sin\triangle_{-n-1} \cos(\triangle_{-n-1}+\pi\alpha)\right].
\label{reshall}
\end{equation}
The result shows that the Hall resistivity is proportional to
the density of vortices and depends on their vorticity via
trigonometrical functions. In particular, as a 
self-consistency check, the Hall resistivity (\ref{reshall}) 
vanishes for $\alpha$ an integer. The Hall resistivity also vanishes
whenever $\triangle_{-n}=-\triangle_{-n-1}$ modulo $\pi$.

\section{Discussion of the results}
The single-particle density of states  $\rho_\alpha$ 
induced by the AB potential was calculated.
It was shown that  $\rho_\alpha$ is a symmetric and
periodic function of the flux.
Existence of the resonance in the AB potential was proven
and the phase-shift flip was discussed.
In addition to the flux, the number of bound states for a 
nonsingular flux tube was also shown to
depend on the energy of the magnetic field.
The Hall resistivity in the dilute vortex limit was calculated.

Our results are not only of academic but also of practical 
interest \cite{BKP} thanks to the recent developments 
in the fabrication of  microstructures and in mesoscopic physics 
(see \cite{WW} for a recent review). In particular,
challenging is 
observation of the resonance and the Hall resistivity.
They occur only in the case if a
bound state is present or, in the latter case, if 
the phase-shift flip occurs.
It will be interesting to consider an application
of our results in the set-up (see Fig. \ref{figrssc}) 
proposed by Rammer and Shelankov
\cite{RS}, especially because
recent measurements on yttrium-barium-copper oxide (YBCO) delta rings 
with three grain-boundary
 Josephson junction \cite{Kir},
reported the observation of vortices that curry a flux 
$\alpha=1/4$ which is
{\em smaller} than the standard flux quantum $h c/2e$ 
(corresponding to $\alpha=1/2$) in the superconductor.
Therefore, when the  high-T${}_c$ YBCO film is used
as a gate  on top of the heterostructure containing 2DEG,
the resonance is at some finite energy and 
could in principle be observed. The same experimental set-up is also
promising for detecting the Hall effect for the random distribution
of vortices.

Our results have been presented in the mathematical 
language of self-adjoint extensions. A self-adjoint extension
is actually the $R\rightarrow 0$ limit, where $R$ is the radius of
a flux tube. Experimentally, infinitely thin means nothing but
that the radius of the flux tube is negligibly small when compared
to any other length, such as a wavelength of particles, in the system.
Therefore, this is the regime in which our results can be applied.
Parameters $\triangle_{-n,-n-1}$ of self-adjoint extensions
are then determined by  bound state energies in the $l=-n$ 
and $l=-n-1$ channels.

I should like to thank A. Comtet, Y. Georgelin, S. Ouvry, and J. Stern
for many useful and stimulating discussions.


\begin{thebibliography}{99}
\bibitem{AB}
W. Ehrenberg and R. E. Siday, {\em Proc. Phys. Soc.} {\bf 62B}, 8 (1949);
Y. Aharonov and D. Bohm, {\em Phys. Rev.}\ {\bf 115}, 485 (1959);
W. C. Henneberger, {\em Phys. Rev.}\  {\bf A22}, 1383 (1980).
\bibitem{AW}M. Alford and F. Wilczek, {\em Phys. Rev. Lett.}\ 
{\bf 62}, 1071 (1989).
\bibitem{F}
J. M. Lifschitz, {\em Usp. Matem. Nauk} {\bf 7}, 170 (1952);
M. G. Krein, {\em Matem. Sbornik} {\bf 33}, 597 (1953);
J. Friedel, {\em Nuovo Cimento Suppl.} {\bf 7}, 287 (1958);
J. S. Faulkner, {\em J. Phys. Solid State Phys.}\ {\bf 10}, 4661 
(1977).
\bibitem{AMB}Recently we have shown that the Krein-Friedel formula
can be applied to the Maxwell equations as well. See
A. Moroz, {\em Phys. Rev.}  {\bf B51}, 2068 (1995).   
\bibitem{AC}Y. Aharonov and A. Casher, {\em Phys. Rev.}\ {\bf A19}, 
2461 (1979).
\bibitem{AM20}A. Moroz, Report IPNO/TH 94-20 (hep-th/9404104).
\bibitem{AM30}A. Moroz, {\em Phys. Rev.}\ {\bf A53}, 669 (1996).
\bibitem{CGO}A. Comtet, Y. Georgelin, and S. Ouvry, {\em 
J. Phys. A: Math. Gen.}\ 
{\bf 22}, 3917 (1989).
\bibitem{NS}A. J. Niemi and G. W. Semenoff, {\em Phys. Rev. Lett.}\ {\bf 51}, 
2077 (1983); T. Jaroszewicz, {\em Phys. Rev.}\  {\bf D34}, 3128 (1986).
\bibitem{AM89}A. Moroz, {\em Phys. Lett.}\ B {\bf 358}, 305 (1995);
Report IPNO/TH 94-89.    
\bibitem{CMO1}A. Comtet, A. Moroz, and S. Ouvry,
{\em Phys.\ Rev.\ Lett.}\ {\bf 74}, 828 (1995).
\bibitem{RS}J. Rammer and A. L. Shelankov, {\em Phys. Rev.}\  
{\bf B36}, 3135 (1987).
\bibitem{BKP}S. J. Bending, K. von Klitzing, and K. Ploog,
{\em Phys. Rev. Lett.}\ {\bf 65}, 1060 (1990).
\bibitem{OO}C. R. Hagen, {\em Phys. Rev. Lett.}\ {\bf 64}, 2347 (1990).
\bibitem{Land}V. B. Berestetskii, E. M. Lifshitz, and
L. P. Pitaevskii, {\em Quantum Electrodynamics}, 2nd. ed.
(Pergamon Press, New York, 1982) \& 116-118.
\bibitem{R}S. N. M. Ruijsenaars, {\em Ann. Phys.}\ {\bf 146}, 1 (1983).
\bibitem{RR}R. D. Richtmayer, {\em Principles of Advanced Mathematical Physics}
(Springer, New York, 1978), vol. 1, Chap. 10.15.
\bibitem{Hag}C. R. Hagen, {\em Phys. Rev. Lett.}\ {\bf 64}, 503 (1990).
\bibitem{La}L. D. Landau, 
{\em Quantum Mechanics}, 3rd ed. (Pergamon Press, Oxford, 1977).
\bibitem{RS2}M. Reed and B. Simon, {\em Fourier Analysis, Self-Adjointness}
(Academic Press, New York, 1975).
\bibitem{BV}M. Bordag and S. Voropaev, {\em J. Phys. A: Math. Gen.}\ {\bf 26}, 
7637 (1993).
We use $e=-|e|$ which leads to $l\rightarrow -l$ when compared to their
notation.
\bibitem{AS}M. Abramowitch and I. A. Stegun, {\em Handbook of Mathematical
Functions} (Dover Publ., 1973).
\bibitem{BF}F. A. Berezin and L. D. Faddeev, {\em Soviet. Math. Dokl.}
{\bf 2}, 372 (1961); C. Manuel and R. Tarrach, {\em
Phys. Lett.}\  {\bf B328}, 
113 (1994).
\bibitem{BV1}We have found that some conclusions of \cite{BV} are not valid:
their coefficients $\alpha_i$ might be negative when they claim them to be
positive; there is an error in calculating the number of bound states. 
\bibitem{NH}M. Nielsen and P. Hedegaard, (cond-mat/9404053).
We disagree, however, with their calculation of the conventional 
phase shifts in the AB potential.
\bibitem{WW}S. Washburn and R. A. Webb, {\em
Rep. Prog. Phys.} {\bf 55}, 1311 (1993).
\bibitem{Kir}C. C. Tsuei et al., 
 {\em Phys. Rev. Lett.}\ {\bf 73}, 593 (1994);
J. R. Kirtley et al., {\em Nature} {\bf 373}, 225 (1995).
\end{thebibliography}
\end{document}